# Machine Unlearning: its nature, scope, and importance for a "delete culture"

Luciano Floridi [1,2]

[1] Oxford Internet Institute, University of Oxford, 1 St Giles', Oxford, OX1 3JS, UK
[2] Department of Legal Studies, University of Bologna, Via Zamboni, 27/29, 40126, Bologna, IT

*Email of correspondence author: luciano.floridi@oii.ox.ac.uk

**Abstract**

The article explores the cultural shift from recording to deleting information in the digital age and its implications on privacy, intellectual property (IP), and Large Language Models like ChatGPT. It begins by defining a *delete culture* where information, in principle legal, is made unavailable or inaccessible because unacceptable or undesirable, especially but not only due to its potential to infringe on privacy or IP. Then it focuses on two strategies in this context: *deleting*, to make information unavailable; and *blocking*, to make it inaccessible. The article argues that both strategies have significant implications, particularly for machine learning (ML) models where information is not easily made unavailable. However, the emerging research area of *Machine Unlearning* (MU) is highlighted as a potential solution. MU, still in its infancy, seeks to remove specific data points from ML models, effectively making them 'forget' completely specific information. If successful, MU could provide a feasible means to manage the overabundance of information and ensure a better protection of privacy and IP. However, potential ethical risks, such as misuse, overuse, and underuse of MU, should be systematically studied to devise appropriate policies.

**Keywords**: ChatGPT, Delete Culture, Machine Learning, Machine Unlearning, Right to be Forgotten.

To introduce this article, it is useful to spell out four premises, even briefly. None of them should come as a novelty.

First, humanity has moved from a millennial culture of *recording* information (I use the term generically, to refer to any content in any medium) to a new culture of *deleting* it. For most of human history, information was analogue and transient by default, because it was only oral, or because oral information had to be encoded with some effort, through physical traces. Until recently, it was difficult to collect and fix information, and ensure that it would survive, whether carved in stone, impressed in the clay of a tablet, hand-written on papyrus or parchment, printed on paper, etched onto a vinyl disc, or recorded on a tape. It was a negentropic struggle, and the crucial decision concerned what to *collect, record, save*, and *transmit* to future generations. Anyone who still remembers loading a film into a camera to take photographs, or recording a VHS will find this familiar. Today, most information is born digital and hence comes as archivable, reproducible, and modifiable records. Analogue information needed to be recorded; most digital information is already available as recorded. So, it accumulates by default, any information society depends on this accumulation and management, and the crucial decision concerns what to *remove* or *delete*—in the general sense of the term, the one that applies to a deleted digital file, which may still be recoverable. Anyone has this experience with the accumulation of pictures in a smartphone or emails in an Inbox. The epochal shift in the ontology of information and hence from scarcity to overabundance has transformed privacy into a defining problem of the digital age. And keeps exacerbating, among other things, quality issues (Floridi 2013; Floridi and Illari 2014) concerning mis- and disinformation, the difficulty of distinguishing the signals from the noise, and the increasing importance of trust, reliability and provenance[1] of information to assess its value and relevance (here blockchain may play a significant role). The problem is no longer too little information but too much.

---

[1] "Provenance" is a technical term comon in art history to refer to the ownership history of an artwork, from when it was first created to its current status, e.g. its presence in a collection or museum. I use it to refer to what is known as "data lineage" in ML: the diachronic tracking of data (information) movement, from the source/producer, through forms of persistence and transformations, to the ultimate consumption.

Second: our newborn deletion culture concerns what information *can*, and if so, *ought* to be made *unavailable* in principle (completely removed or irrecoverably deleted), or at least *inaccessible* in practice—that is, still available yet not obtainable—and in *what circumstances*, that is, when, where, to whom, how long for, for what purposes, and so forth. Note the order of "can" and "ought". It is a classic and uncontroversial principle, in ethics and jurisprudence, that "ought" implies "can", in the logical sense of "if *a* ought to do *p*, then *p* can be done, that is, it must be possible for *a* to do *p*". No ethical or legal request should ever impose an action (ought) if that action is impossible (cannot). So, any solution in managing our delete culture must be feasible before being imposed as an obligation. This may seem obvious, but some legal requests to do things differently (one ought to do *p* instead of *q*) may, in fact, amount to stopping something from happening altogether (because one cannot *p*, then one must stop doing *q*). The example of the Italian *de facto* block of ChatGPT in Italy is a case in question (see below).

Third: in a delete culture, content that is not *illegal in itself*—for example, content that is not forbidden because it is pornographic, violent, etc., this clarification is essential to understand the examples below and avoid conflating a delete culture with a *censorship culture*—can still be removed (made unavailable or inaccessible) usually because its availability/accessibility infringe *intellectual property* (IP) or *privacy*. There are other, important reasons that I shall only briefly mention but not discuss in this article, such as *secrecy* and *security*.[2] Encrypted information, for example, is available yet inaccessible to some people because of such reasons. In a delete culture, in which information grows by default, encryption, and other "inaccessibility" practices, like password protection, become increasingly common. However, for the sake of simplicity, I shall ignore these topics in the following pages and limit myself to the two examples of privacy and IP. They are sufficiently pressing and illustrative.

Fourth: in a delete culture that has just begun to develop—and that should not be confused with a *cancel culture*, although the shift in the cultural focus from recording to deleting might have influenced the mentality behind it—the risk is to keep using some conceptual tools that were refined by a recording culture to cope with problems like infringements of privacy or intellectual property, when in fact new approaches may be needed (note that the cancel culture runs the same risk). I shall return to this topic below.

---

[2] Machine Unlearning could be useful to mitigate poisoning attacks on machine learning models (Liu et al. 2022; Marchant, Rubinstein, and Alfeld 2022) because if a machine learning model is compromised by the injection of some pollution data into its dataset, it becomes necessary to remove such data from the dataset and retrain it.

Given these premises, and because there are two senses in which information can be removed, we saw that there are currently two main strategies that can be pursued when some digital information can and ought to be removed: delete it to make it unavailable, or block it to make it inaccessible. Remarkably, this distinction seems disregarded in the literature about *Machine Unlearning* (a new research field in AI, to be explained presently), which often misrepresents the right to be forgotten for a right to have one's own private information deleted instead of blocked. Yet, the distinction is crucial, conceptually and technically, and because the two strategies come with very different costs and degrees of implementability and flexibility. Blocking information is much cheaper and easier than removing it, and it can be reversed much more easily. Here are two examples that are now textbook cases in any class on digital ethics. I present them in logical and not chronological order.

Following the Cambridge Analytica scandal in 2018, the UK Information Commissioner Office required Cambridge Analytica to *delete* the data it held.[3] This is a case of making the information *unavailable* permanently (of course, people may not comply, by not really deleting the data or keeping copies, but this is irrelevant here). Note the combination of the three variables. In this case, we have: unavailability of the information + inaccessibility of the same information + in *all* circumstances.

The second example concerns the landmark decision by the Court of Justice of the European Union (CJEU) in 2014 about Google Spain v AEPD and Mario Costeja González. Following this so-called "right to be forgotten" case (Floridi 2015), Google was required not to include information about González's financial history in the results of its search engine.[4] The solution has kept evolving in the following years, both in the EU and internationally. However, the essential point remains that the relevant information is *not* deleted (it is still *available*); it is only made *inaccessible* in the versions of the search engine available in the region involved, for example in the EU. Thus, one can still access it in other regions or in the EU through a VPN. In this case, the combination of the three variables is: availability of the information + inaccessibility of the same information + in *some* circumstances, for example, in the EU. The same approach can be seen at work when YouTube videos are not accessible in some countries, such as Germany or Turkey, because of local legislation.

---

[3] See https://ico.org.uk/media/action-weve-taken/2618383/20201002_ico-o-ed-l-rtl-0181_to-julian-knight-mp.pdf

[4] Disclosure: I was a memberof Google's Advisory Board.

The "available but inaccessible" solution is a compromise. Because it is usually (not always, see the encryption case, for example, or the use of robot.txt files that prevent web crawlers) based on IP address geolocation and hence on the (in this case outdated) Westphalian principle of the territoriality of the law, it can easily be made ineffective using a VPN, or any browser with a free VPN service integrated by default. The law is formally applied locally, but, substantially, it relies only on some friction, introduced by practical inertia and a minor technical (and sometimes financial) barrier. Anyone who is determined to access the information can do so easily. A similar reasoning applies to "secret" information: anything that remains available risks becoming accessible, no matter what protection may be in place, so the whole strategy is focused on how difficult it can be. However, "blocking" is inexpensive and easy to implement, it is more flexible (different circumstances receive different "inaccessibility" treatment), and it can be reversed, as inexpensively, easily, and flexibly as it is put in place. I shall return to this point at the end of the article.

The two alternatives, *remove* (unavailability) and *block* (inaccessibility), have dominated our delete culture for decades. Thus, they have also shaped the debate on Large Language Models in general, and especially the extremely popular ChatGPT.[5] What can be done if the information provided by ChatGPT is unacceptable, for example, because of intellectual property or privacy considerations? Recall that "the information *ought* to be removed" requires that "the information *can* be removed", and that, in a delete culture, it is essential to develop tools that can facilitate the efficient and effective management of the overabundance of information, including the unwanted kind. Since ChatGPT (and all other chatbots like it) are not databases—despite misleading remarks often found in some ill-informed, legal, and political discussions about them—it *seems* (more on this below) that their information cannot be made unavailable but only inaccessible, short of making the whole service illegal. The system has learnt from the data, but does not hold the data, so there is no data to erase, like with Cambridge Analytica. Apparently, the only solution is to block the information in question, that is, to make the output of the system inaccessible, like in the case of Google search engine and the right to be forgotten. This is done through IP address geolocation, and it is precisely what happened when the Italian Data Protection Authority (Garante per la protezione dei dati personali) required OpenAI to suspend, with immediate effect, the processing of personal data of Italian users. OpenAI complied and made ChatGPT no longer available to Italy-based users. Of course, anyone in Italy could keep using the service through a VPN. Later, the service became available again, after some minor adjustments were made (including an age verification mechanism, which, no matter how loosely or strictly implemented, is in line with the inaccessibility principle in some circumstances, in this case, the user's age and location).

---

[5] Similar considerations apply to Text-to-Image (TTI) systems like Dall-E, but there are technical differences and different implications, so I shall not discuss them in this article.

The question at this point is: when it comes to machine learning (ML), once a model has been trained,[6] if there is a problem concerning privacy or intellectual property (or safety, security, secrecy etc.), is there any other solution, apart from (a) deleting the model completely; (b) removing the unwanted data from the training set and then entirely retrain the model from scratch ((Thudi, 2022) calls this *exact unlearning*, but it seems a misnomer, as it is really *re-learning*); or (c) "blocking" the information that should not be obtainable, once the model has learnt to produce it? The question is pressing because (a) can easily be too radical; (b) can be very expensive, financially, computationally and in terms of the time required; and (c) can be utterly ineffective,[7] so it depends on how crucial it is to make the information unavailable rather than just inaccessible. Luckily, the answer is a tentative yes, for there is a fourth alternative, already mentioned above: *Machine Unlearning* (MU; I shall use this label to refer preferably only to this kind of partial or selective, and potentially incremental "amnesia"; it is also called *approximate unlearning* in (Thudi, 2022).

---

[6] I prefer not to include in the list differential privacy achieved by adding noise (and related loss in performance) because it is too specific and works at a different level of abstraction (Sekhari et al. 2021), and also because it refers ony to privacy issues and is not easily extendable to other ethical or legal problems. However, it may be added, if necessary, as a kind of approximate unlearning, as done by (Zhang, Nakamura, et al. 2023).
[7] See for example the many "jailbreakss that cant trick a chatbot, hre is an example: https://tcrn.ch/3Lifc9N

MU is a recent area of research in AI (Nguyen et al. 2022; Zhang, Nakamura, et al. 2023).[8] (Cao and Yang 2015) is the influential paper that introduced the idea and the terminology (for a recent introduction, see (Mercuri et al. 2022)). Conceptually, MU is not difficult to understand. In ML, models are trained on datasets containing examples of the problem that the model is intended to solve. During training, the model learns to recognise patterns in the available data and make predictions based on those patterns or associations, outputting the corresponding information. Depending on the dataset—which, for example, may include sensitive data—the model may learn patterns or associations that deliver information that is unacceptable or undesirable, for epistemological (e.g., because it is outdated, inaccurate, incorrect, or misleading), ethical (e.g., because it is biased, unfair, or discriminatory), or legal (e.g., because it breaches privacy or IP rights) reasons. MU addresses these issues by removing from the model all the traces of some data points (a sort of selective amnesia), without affecting the performance, so that the model no longer delivers the corresponding, problematic information. Thus, MU can be understood as reversing the process that led a model to learn how to provide some specific information. It trains the ML model again, to deliver information with a portion of the data forgotten, to minimize the costs associated with complete retraining from scratch.

For example, suppose one asks ChatGPT some financial information about Mario Costeja González's bankruptcy. Assuming that ChatGPT can answer, it may still block the answer, that is, make it inaccessible, explaining that the information is not accessible because of the decision taken by the CJEU. This is what could happen (or at least should happen) now in the EU. Using a very anthropomorphic language, ChatGPT "knows" the answer but "refuses" to provide it. This would be a compromise, as we saw above, and it could be made pointless through technological solutions, such as prompts devised to deceive ChatGPT into delivering the information. It would be preferable if ChatGPT did not know the answer at all. MU could achieve this. So, in the near future, the CJEU could require a company to ensure that a chatbot *unlearns* some specific information, making it unavailable in principle, not just inaccessible in practice. In the future, users may ask search engines to comply with their "right to be forgotten" and ML systems with their "right to be unlearnt".

---

[8] The field is sufficiently young still to allow the editing of a collection of papers, see for example: https://github.com/jjbrophy47/machine_unlearning#Before-2017.

MU is still in its infancy and not devoid of difficulties, including computational efficiency and scalability, technical reliability and certifiability (Bourtoule et al. 2021; Sekhari et al. 2021)) and management issues (MU can also be subject to malicious attacks (Chen et al. 2021). It may require provability and verifiability (Eisenhofer et al. 2022; Sommer et al. 2022)). It also seems that MU works better the more an ML model is designed with a later MU application in mind, in a sort of "pre-unlearning" strategy: unlearning data A instead of data B is easier if you structure the learning process in such a way to discriminate between A and B (see for example the Sharded, Isolated, Sliced, and Aggregated (SISA) approach (Bourtoule et al. 2021)), yet this does not come without costs (Koch and Soll 2023). Indeed, some argue that MU is "well defined only at the algorithmic level" (Thudi et al. 2022). Certainly, the recent, international debate about the decision of the Italian Data Protection Authority to block ChatGPT saw almost no references to MU as a possible solution.[9]

However, there are already different approaches to how MU can be achieved (Aldaghri, Mahdavifar, and Beirami 2021; Warnecke et al. 2021; Mahadevan and Mathioudakis 2021; Tarun et al. 2023; Zhang, Nakamura, et al. 2023) and comparative assessment of MU methods (Zhang, Pan, et al. 2023). And MU is a promising way of addressing growing needs in a delete culture. So, the field is likely to gain momentum as legal and social pressures increase, especially because of the AI Act and the (proposal for an) AI Liability Directive. The time when an AI system uses MU to train an ML model to remove specific data points and unlearn some specific patterns so that it does not deliver the corresponding information may not be too distant. Supervised MU, even if it were to require human input or oversight, could significantly improve the quality of the information provided by ML models, not only epistemologically, ethically, and legally, but also economically, e.g., in terms of fine-tuning, efficiency, and effectiveness. But this is not the place where to review such scenarios. What matters here is that, *if* MU thrives—and the "if" is primarily technical and scientific—our new delete culture will soon be able to offer unlearning as one more way to make some information not just inaccessible but also unavailable. For example, in the future, the request to remove personal information of EU citizens from ChatGPT would not be satisfied by a compromise, that is, by making it inaccessible, but by a feasible operation of unlearning, which respects in full both the spirit of the law and the "ought implies can" principle. Future versions of ChatGPT may offer the option of asking to unlearn our private information. This will be a much more adequate solution for a twenty-first-century problem.

---

[9] For an exception see for example https://www.jdsupra.com/legalnews/an-elephant-never-forgets-and-neither-7307484/ which is, however, critical about the viability of MU.

As always, MU faces not only technical challenges but also ethical risks. They should be studied systematically, also in view of designing policies that can avoid or minimise them. Here, let me conclude this article by outlining three kinds. In general, some pressing difficulties, discussed in the context of the implementation of the right to be forgotten, reappear in different formats but with unmitigated pressure.

Let me start with misuse. MU takes as its input a trained model that provides some undesirable or unacceptable information and produces as its output a revised model that no longer provides that information. The problem is to decide what is undesirable or unacceptable. In the hands of malicious or unscrupulous agents, from autocrats to criminals, MU could become a powerful tool for censorship, misinformation, propaganda, vandalism, discrimination, cyber-attacks, or new forms of ransomware, making acceptable and desirable information no longer available. Less stark cases (think of some information omission as a case of PR) will provide more challenging questions, linked to the debate about the ethical decisions steering and formatting the training of chatbots like ChatGPT. In a predictable world in which too much information increasingly requires ML tolls to be efficiently and effectively managed, the misuse of MU could transform the "editing" away of information into a massive *damnatio memoriae* of 1984 proportions.

Next, consider overuse. Counterintuitively, the development of successful MU methods could also invite good actors to take more risks, not fewer, and more lightly, when training their ML models. If models can unlearn quickly, easily, and inexpensively, then "better ask forgiveness (that is, unlearn later) than permission (not learning in the first place)" could become a default approach. This seems to have already happened with social media and the "inaccessibility" solution for the right to be forgotten: collect first, block later, if required. MU could support a cynical cost-benefit analysis.

Finally, MU could be underused. I mentioned above that removing and blocking are two different approaches, which come with different costs, difficulties, and flexibility in their implementation. Blocking information, a sort of "machine silencing", may turn out to be the biggest competitor of any MU strategy, unless MU ends up being required by law, for cases that need much more than just making some information inaccessible in some circumstances. Recall that Cambridge Analytica was required to delete its information, not just to stop using it or to make it inaccessible. In the future, the debate may be whether unlearning or blocking will be the right approach to some information too sensitive to be left available.

It has taken millennia to develop a culture of recording. It will not take decades to refine one of deleting, but it seems that MU will likely be part of such development. It is definitely worth investing in its development and study.

**Acknowledgements**

This article would have been impossible without the input from Rita Cucchiara, who called my attention to the importance of the phenomenon, and Emmie Hine, Claudio Novelli, Samuele Poppi, and Mariarosaria Taddeo for our conversations and their feedback.